\documentclass[showpacs,aps,prc,longbibliography,showkeys,superscriptaddress,twocolumn,nofootinbib]{revtex4-1}

\usepackage[colorlinks, urlcolor=blue,linkcolor=blue, anchorcolor=blue, citecolor=blue]{hyperref}
\usepackage[english]{babel}
\usepackage[utf8]{inputenc}
\usepackage{multirow}
\usepackage{array}
\usepackage{amsthm}
\usepackage{mathtools}
\usepackage{physics}
\usepackage{xcolor}
\usepackage{graphicx}
\usepackage{bm}
\usepackage{soul}
\usepackage{adjustbox}
\usepackage{placeins}
\usepackage[T1]{fontenc}
\usepackage{lipsum}
\usepackage{csquotes}
\usepackage{float}
\usepackage{booktabs}
\usepackage{lineno}

\graphicspath{{figs/}}

\begin{document}


\author{Lipei Du}
\affiliation{Department of Physics, University of California, Berkeley CA 94720}
\affiliation{Nuclear Science Division, Lawrence Berkeley National Laboratory, Berkeley CA 94720}

\date{\today}

\title{Physical interpretation and limits of photon--dilepton radial-flow tomography}

\begin{abstract}
Thermal photons and dileptons provide complementary information on the
temperature and collective expansion of matter created in relativistic
heavy-ion collisions.  Their combination has been proposed as an
electromagnetic probe of radial flow: a comparatively flow-insensitive
dilepton invariant-mass inverse slope is used to infer the photon inverse
slope expected without transverse flow and compared with the observed
photon slope.  Here the physical basis and leading limitation of
this construction are examined using a controlled expanding-fireball
calculation.  The dilepton and flow-free photon inverse slopes are tightly
correlated because they respond similarly to changes of the underlying
thermal scale, although the numerical relation depends on the spectral
window and electromagnetic source.  The flow-induced change of the photon
inverse slope correlates more strongly with the photon-emission-weighted
velocity component along the photon momentum than with a bulk-like
radial-flow average. Prethermal source variations can bias the inferred
flow-free photon reference and generate apparent flow even when the direct
photon spectral-flow signal vanishes, while multiple photon and dilepton
windows provide independent response patterns that can help diagnose this
source ambiguity.  These results clarify both what photon--dilepton
radial-flow tomography measures and the source consistency required for its
interpretation.
\end{abstract}

\maketitle

\section{Introduction}

Collective expansion is one of the principal manifestations of the pressure
generated in relativistic heavy-ion collisions and provides essential
information on the equation of state and transport properties of the
quark--gluon plasma (QGP)
\cite{Heinz:2013th,Sorensen:2023zkk}.  Most information on radial flow comes
from final-state hadrons, which reflect the accumulated expansion after the
medium has cooled substantially \cite{Shuryak:2014zxa,Arslandok:2023utm}.  Electromagnetic radiation instead escapes
with negligible final-state interaction and is emitted throughout the
evolution
\cite{Peitzmann:2001mz,Gale:2009gc,Rapp:2013nxa,Linnyk:2015rco,
Geurts:2022xmk,Salabura:2020tou}, providing access to the earlier development of collective motion.

The same continuous emission that makes electromagnetic probes valuable also
complicates their interpretation.  Over a finite transverse-momentum window,
the thermal-photon spectrum can be characterized by an exponential
inverse-slope parameter $T_\gamma$, with dimensions of temperature, that
reflects both the temperature history and transverse Doppler effects
\cite{Shen:2013vja,Paquet:2022wgu,Paquet:2023bdx}.
Because the observed spectrum integrates radiation over the full spacetime
evolution, this inverse slope cannot in general be interpreted as the blue
shift of a single local temperature.  In particular,
transverse flow can increase or decrease the fitted photon inverse slope
depending on the momentum window \cite{Paquet:2022wgu}.  Intermediate-mass
dileptons provide complementary information
\cite{Shuryak:1978ij,Kajantie:1981wg,McLerran:1984ay,Kajantie:1986dh}.  Their invariant-mass spectrum can likewise be
characterized by an inverse-slope parameter $T_{\ell\bar\ell}$, while the
invariant mass is much less sensitive to radial flow
\cite{Rapp:2014hha,HADES:2019auv,Churchill:2023vpt,Churchill:2023zkk,Du:2024pbd}.  The dilepton
inverse slope therefore primarily probes an emission-weighted temperature
scale whose relation to the thermal history depends on the cooling evolution
and source composition
\cite{Massen:2024pnj,Vujanovic:2013jpa,Wu:2024pba,Du:2026aux}.
Dilepton measurements have established the experimental basis for
using these spectra to study the evolving medium
\cite{STAR:2013pwb,STAR:2015tnn,STAR:2024bpc,Bailhache:2025kwa}.  The photon and dilepton inverse
slopes thus encode different aspects of the same evolving electromagnetic
source.

This complementarity motivated a photon--dilepton construction for
extracting radial flow from electromagnetic spectra
\cite{Du:2024pbd,Du:2025dot}.  The method proceeds in two conceptually distinct steps.
First, the approximately linear relation between the dilepton inverse slope
$T_{\ell\bar\ell}$ and the flow-free photon inverse slope $T_{\gamma0}$ is
used to infer the photon inverse slope that would be obtained from the same
source in the absence of transverse flow.  Second, this inferred reference
is compared with the observed photon inverse slope $T_\gamma$ through a
Doppler-motivated relation to define an effective radial-flow signal.  The two steps raise different questions.  The first concerns why the
$T_{\ell\bar\ell}$--$T_{\gamma0}$ relation is so nearly linear, how its
numerical form depends on the spectral selection, how stable it remains
under changes of the thermal evolution, and how reliably it can be
transferred across electromagnetic source compositions.  The second
concerns what average of the spacetime-dependent radial velocity field is
represented by the effective flow defined by comparing the observed photon
inverse slope $T_\gamma$ with its flow-free reference $T_{\gamma0}$.

Earlier multimessenger
calculations at Beam Energy Scan energies explored
photon and dilepton spectral temperatures and the possibility of combining
the two probes to access radial flow \cite{Du:2024pbd}.  Subsequent
hydrodynamic calculations have established both a tight
$T_{\ell\bar\ell}$--$T_{\gamma0}$ correlation and a strong correlation
between the reconstructed effective flow and the pre-freezeout radial
expansion \cite{Du:2025dot}.  The remaining questions are how these two
relations should be interpreted physically, how robust they are under
controlled variations of the evolution, and how they are modified by
prethermal radiation \cite{Churchill:2020uvk,Gale:2021emg,Wu:2024pba}, which was not
included in those calculations.

These questions are examined in a controlled calculation that varies the
temperature history, radial flow, and properties of the prethermal source
independently.  The analysis first considers the
$T_{\ell\bar\ell}$--$T_{\gamma0}$ thermometer relation, including the
origin of its near-linearity and its dependence on the photon momentum
window.  The directly calculated $T_{\gamma0}$ is then used to bypass this
inference step and identify the collective-flow quantity most closely
associated with the flow-induced change of the photon inverse slope.
Finally, the full reconstruction is restored to test how changes in
electromagnetic source composition bias the inferred flow and whether
additional photon and dilepton spectral windows can diagnose that
ambiguity.  This separation clarifies both the physical interpretation and
a leading controlled limitation of photon--dilepton radial-flow tomography.

\section{Controlled fireball and spectral construction}

To isolate the physics entering photon--dilepton radial-flow tomography, the
analysis uses a cylindrically symmetric, boost-invariant fireball
\cite{Paquet:2022wgu,Du:2026aux} rather than a full
hydrodynamic evolution.  This reduced setup keeps the temperature history,
transverse flow, and electromagnetic emission weighting explicit, allowing
their roles in the spectral response to be varied independently.  In a full
hydrodynamic evolution these ingredients evolve together and are therefore
more difficult to vary separately, obscuring how each one influences the
photon and dilepton observables.

The onset of the thermal stage is set to $\tau_0=1~\mathrm{fm}/c$.
For $\tau\geq\tau_0$, the temperature field is parametrized by a Gaussian
transverse profile with Bjorken-like cooling \cite{Bjorken:1982qr},
\begin{equation}
T(\tau,r)
=
T_{0,\mathrm{peak}}
\left(\frac{\tau_0}{\tau}\right)^{c_s^2}
\exp\left(-\frac{r^2}{2\sigma_0^2}\right),
\label{eq:temperature}
\end{equation}
where $T_{0,\mathrm{peak}}$ is the peak temperature at $\tau_0$ and
$r=0$, $\sigma_0=3.01~\mathrm{fm}$ is the transverse width, and
$c_s^2=1/3$ is the squared speed of sound for the conformal equation of
state used here.  The transverse width is held fixed in the primary scans below.

Because the later analysis tests the sensitivity of the tomography to
radiation and radial motion before the thermal stage, the reduced description
is extended back to $\tau_{\rm init}=0.1~\mathrm{fm}/c$.  In the prethermal
interval, $\tau_{\rm init}\leq\tau<\tau_0$, the temperature field of
Eq.~\eqref{eq:temperature} is continued backward in proper time while keeping
the same transverse profile.  This backward-continued scale is denoted by
$T_{\rm pre}^{\rm eff}$.  Its central value is capped at
$0.6~\mathrm{GeV}$ before the Gaussian transverse profile is applied.  The
superscript ``eff'' emphasizes that, before the onset of the thermal stage,
this quantity is used as the energy scale for evaluating electromagnetic
emission and should not be interpreted as an equilibrium thermodynamic
temperature.

For the radial expansion, the approximate central-fireball solution of
Ref.~\cite{Paquet:2022wgu} is adapted by allowing the onset time of radial
acceleration to be varied.  For a generic onset time $\tau_a$, the transverse
four-velocity is defined as
\begin{equation}
u_{\perp,\tau_a}^{\rm base}(\tau,r)
=
\frac{r}{\sigma_0}
\frac{
\tau-\tau_a(\tau/\tau_a)^{c_s^2}
}{
(1-c_s^2)\sigma_0
\left[1+\tau^2/(2\sigma_0^2)\right]
},
\qquad \tau\geq\tau_a,
\label{eq:base-flowfield}
\end{equation}
and set $u_{\perp,\tau_a}^{\rm base}=0$ for $\tau<\tau_a$.  The original
thermal-stage solution corresponds to $\tau_a=\tau_0$.  The minimal
reference prescription is therefore
\begin{equation}
u_\perp^{(0)}(\tau,r)
=
\kappa_r
u_{\perp,\tau_0}^{\rm base}(\tau,r),
\label{eq:flowfield}
\end{equation}
which vanishes before $\tau_0$ and will be referred to below as the
zero-before-$\tau_0$ velocity history.  This choice avoids introducing a separate model for prethermal radial
motion and provides a minimal reference against which earlier flow
development can be tested.  The sensitivity to such an earlier buildup is
examined through the alternative history below.  The factor $\kappa_r$ is a
control parameter introduced here and is not part of the original
approximate solution~\cite{Paquet:2022wgu}.  Varying it at fixed temperature history does not
represent a self-consistent hydrodynamic evolution; rather, it deliberately
separates the spectral response to transverse flow from that to cooling.

To test sensitivity to an earlier buildup of radial motion, an alternative
velocity history is constructed in which radial flow begins developing at
$\tau_{\rm init}$, before the onset of the thermal stage at $\tau_0$.
This history is obtained by interpolating between two versions of the same
analytic flow solution that differ only in the assumed onset time of radial
acceleration.  The first begins at $\tau_0$ and corresponds to the reference
zero-before-$\tau_0$ history, while the second begins at
$\tau_{\rm init}$ and therefore develops radial motion already during the
prethermal interval.  Since $u_\perp=\sinh\rho_\perp$, with $\rho_\perp$
the transverse rapidity, the interpolation is performed at the level of
$\rho_\perp$.  Defining
\begin{equation}
\rho_{\perp,\tau_a}^{\rm base}(\tau,r)
=
\operatorname{arsinh}
u_{\perp,\tau_a}^{\rm base}(\tau,r),
\end{equation}
the interpolated history is
\begin{equation}
u_\perp^{(f)}(\tau,r)
=
\kappa_r\sinh\Big[
(1-f)\rho_{\perp,\tau_0}^{\rm base}(\tau,r)
+
f\rho_{\perp,\tau_{\rm init}}^{\rm base}(\tau,r)
\Big].
\label{eq:continuous-flow}
\end{equation}
Thus $f=0$ exactly reproduces the zero-before-$\tau_0$ history, whereas
$f=1$ gives the same analytic flow solution with radial motion beginning at
$\tau_{\rm init}$.  The calculations below use $f=0.25$ as a moderate
controlled variation, corresponding to a history lying $25\%$ of the way in transverse
rapidity from the $\tau_0$-onset solution toward the
$\tau_{\rm init}$-onset solution.  This produces nonzero radial motion
during the prethermal interval and modifies the subsequent thermal-stage
flow history as well.  The value $f=0.25$ is not assigned the meaning of an
inferred physical fraction.

For either velocity history, the local radial three-velocity and Lorentz
factor are
\begin{equation}
v_r(\tau,r)
=
\frac{u_\perp(\tau,r)}
{\sqrt{1+u_\perp^2(\tau,r)}} ,
\qquad
\gamma=\sqrt{1+u_\perp^2}.
\label{eq:local-radial-velocity}
\end{equation}

With the temperature and velocity histories specified, the photon and
dilepton spectra are constructed using the same electromagnetic-rate framework
\cite{Gale:1987ki,Laine:2013vma,Ghisoiu:2014mha,Ghiglieri:2014kma,Ghiglieri:2021vcq,Jackson:2019yao,Arnold:2001ms}.  For the thermal stage, the midrapidity
contributions are
\begin{align}
\left.
\frac{1}{2\pi p_T^\gamma}
\frac{d^2N_\gamma^{\rm th}}{dp_T^\gamma dy}
\right|_{y=0}
&=
\int d^4x\,
\left.
E^*
\frac{d\Gamma_\gamma}{d^3p^*}
\right|_{E^*=p\cdot u(x)},
\label{eq:photon-spectrum}
\\
\left.
\frac{1}{2\pi M}\frac{dN_{\ell\bar\ell}^{\rm th}}{dMdy}
\right|_{y=0}
&=
\int dp_T^{\ell\bar\ell}\,p_T^{\ell\bar\ell}
\int d^4x\,
\frac{d\Gamma_{\ell\bar\ell}}{d^4q}
\left(M,q\cdot u,T(x)\right).
\label{eq:dilepton-spectrum}
\end{align}
Here $p$ and $q$ denote the photon and dilepton-pair four-momenta,
respectively.
Thermal emission is retained for $T\geq0.18~\mathrm{GeV}$, and the same
lower cutoff is applied to $T_{\rm pre}^{\rm eff}$ for prethermal emission.
The dilepton spectrum is integrated over
$0.2<p_T^{\ell\bar\ell}<4.5~\mathrm{GeV}$.  Because invariant mass is
Lorentz invariant, the momentum-integrated IMR spectrum is not Doppler
shifted in the same manner as the photon transverse-momentum spectrum
\cite{Du:2024pbd,Paquet:2022wgu,Paquet:2015lta}; only a small residual flow
dependence remains from the finite pair-momentum acceptance.

In the prethermal interval, the electromagnetic emission is evaluated with
the same kinematic and spacetime integrations as in
Eqs.~\eqref{eq:photon-spectrum} and \eqref{eq:dilepton-spectrum}, replacing
the thermal temperature by $T_{\rm pre}^{\rm eff}$ and modifying the local
rates to account for a chemically undersaturated quark content.  Motivated
by the expectation that the early medium can be gluon rich and approach
quark chemical equilibrium only gradually
\cite{Kurkela:2018oqw,Kurkela:2018xxd,Coquet:2021lca}, a phenomenological quark
fugacity factor is introduced \cite{Gale:2021emg},
\begin{equation}
\lambda_q(\tau)
=
1-\exp\left[-\ln(10)\frac{\tau}{\tau_{\rm chem}}\right].
\label{eq:quark-fugacity}
\end{equation}
This parametrization, expressed in the collision proper time $\tau$, rises
monotonically toward unity and satisfies
$\lambda_q(0)=0$ and $\lambda_q(\tau_{\rm chem})=0.9$, so
$\tau_{\rm chem}$ controls the timescale over which the quark content
approaches its chemically equilibrated limit without changing the
underlying effective-emission-scale history.  For photons, a $\lambda_q$
weighting of the prethermal rate is adopted, motivated by important
production channels such as $qg\rightarrow q\gamma$,
whose leading dependence in a gluon-rich, quark-undersaturated medium involves
a single quark occupancy.  By contrast, the leading IMR dilepton process
$q\bar q\rightarrow\gamma^*\rightarrow\ell^+\ell^-$ requires both a quark and an antiquark and therefore
motivates a $\lambda_q^2$ weighting \cite{Wu:2024pba}.  These fugacity factors are intended as
a controlled parametrization of the different sensitivity of photons and
dileptons to early quark chemical equilibration~\cite{Kurkela:2018wud,Kurkela:2018vqr,Schlichting:2019abc,Berges:2020fwq}, rather than as a complete
microscopic treatment of the prethermal electromagnetic rates.

To test how the photon--dilepton observables respond to a prethermal
contribution of varying strength, an additional dimensionless factor
$\alpha_{\rm pre}$ is introduced that controls the overall magnitude of the
prethermal emission independently of its chemical evolution.  Denoting the
thermal and prethermal contributions by $Y_{\rm EM}^{\rm th}$ and
$Y_{\rm EM}^{\rm pre}$, respectively, the source mixture is written
schematically as
\begin{equation}
Y_{\rm EM}
=
Y_{\rm EM}^{\rm th}
+
\alpha_{\rm pre}Y_{\rm EM}^{\rm pre}.
\label{eq:source-mixture}
\end{equation}
Here $Y_{\rm EM}$ may represent either an integrated yield or the
corresponding differential spectrum; in practice, the same linear
combination is applied bin by bin to the photon transverse-momentum spectrum
and the dilepton invariant-mass spectrum.  Thus $\alpha_{\rm pre}=0$
corresponds to thermal emission only, while $\alpha_{\rm pre}=1$ gives the
reference thermal-plus-prethermal source used below.  Its role is distinct
from that of $\tau_{\rm chem}$: $\alpha_{\rm pre}$ changes the overall
strength of the prethermal radiation without altering its chemical
evolution, whereas $\tau_{\rm chem}$ changes the time dependence of the
quark content and therefore affects photons and dileptons differently
through their respective powers of $\lambda_q$.  Unless stated otherwise,
calculations including prethermal radiation use $\alpha_{\rm pre}=1$ and
$\tau_{\rm chem}=1~\mathrm{fm}/c$.  The two parameters therefore provide
independent controls over the overall strength and chemical equilibration
of the early electromagnetic source.

The resulting photon and dilepton spectra are characterized by the
finite-window inverse slopes that enter the photon--dilepton tomography.
These spectral scales are extracted by fitting \cite{Shen:2013vja,Rapp:2014hha}
\begin{align}
\ln\left[
\frac{1}{2\pi p_T}
\frac{d^2N_\gamma}{dp_Tdy}
\right]
&=
a_\gamma-\frac{p_T}{T_\gamma},
\label{eq:photon-fit}
\\
\ln\left[
M^{-3/2}
\frac{dN_{\ell\bar\ell}}{dMdy}
\right]
&=
a_{\ell\bar\ell}-\frac{M}{T_{\ell\bar\ell}},
\label{eq:dilepton-fit}
\end{align}
at midrapidity using $0.8<p_T<2~\mathrm{GeV}$ for photons and
$1<M<3~\mathrm{GeV}$ for dileptons unless stated otherwise.  To isolate the
effect of transverse flow on the photon spectrum, the flow-free photon
inverse slope $T_{\gamma0}$ is also calculated using the same
temperature and electromagnetic source history, but with the transverse
velocity field set to zero throughout the photon emission calculation while
all other source ingredients are held fixed.  The quantities $T_\gamma$, $T_{\gamma0}$,
and $T_{\ell\bar\ell}$ are therefore fitted spectral inverse-slope
parameters, with dimensions of temperature, rather than local
thermodynamic temperatures \cite{Du:2024pbd,Du:2025dot}.

The first step of photon--dilepton tomography is to infer the experimentally
inaccessible $T_{\gamma0}$ \cite{Du:2024pbd,Du:2025dot}.  For a specified electromagnetic source and
choice of spectral windows, a family of model calculations is generated and
$T_{\gamma0}$ is determined directly from the corresponding photon spectra
with transverse flow removed.  The resulting pairs of
$T_{\ell\bar\ell}$ and $T_{\gamma0}$ define an approximately linear
thermometer relation,
\begin{equation}
T_{\ell\bar\ell}
=
A+B T_{\gamma0},
\label{eq:thermometer-general}
\end{equation}
where $A$ and $B$ are obtained from the model ensemble for the specified
source prescription and spectral selection.  Once this relation is
established, the dilepton inverse slope of an individual spectrum can be used
to infer
\begin{equation}
\widehat T_{\gamma0}
=
\frac{T_{\ell\bar\ell}-A}{B}.
\label{eq:inferred-photon-reference}
\end{equation}
Thus $T_{\gamma0}$ denotes the directly calculated flow-free photon slope,
whereas $\widehat T_{\gamma0}$ denotes its reconstruction from the
thermometer relation.

The second step compares the photon slope with the flow-free reference
\cite{Du:2025dot}.  For radiation from a thermal fluid element moving parallel
to the observed photon momentum with speed $v_r$, the relativistic Doppler
relation gives
\begin{equation}
T_\gamma
=
T_{\gamma0}
\sqrt{\frac{1+v_r}{1-v_r}}
\qquad\Longrightarrow\qquad
v_r^{\rm eff}
=
\frac{T_\gamma^2-T_{\gamma0}^2}
     {T_\gamma^2+T_{\gamma0}^2}.
\label{eq:veff}
\end{equation}
An expanding fireball does not have a single temperature or velocity, so
the second expression is used to \emph{define} the effective spectral flow
associated with the two finite-window photon slopes.  When the directly
calculated $T_{\gamma0}$ is available,
Eq.~\eqref{eq:veff} defines the direct $v_r^{\rm eff}$.  Using instead the
dilepton-inferred $\widehat T_{\gamma0}$ defines
\begin{equation}
\widehat v_r^{\rm eff}
=
\frac{T_\gamma^2-\widehat T_{\gamma0}^{\,2}}
     {T_\gamma^2+\widehat T_{\gamma0}^{\,2}}.
\label{eq:reconstructed-veff}
\end{equation}
The distinction between the direct and reconstructed quantities allows the
thermometer inference and the Doppler response to be tested separately before
they are combined in the full reconstruction.

\section{Results and discussion}

\subsection{Thermometer relation for the flow-free photon slope}

The first test concerns the initial step of photon--dilepton radial-flow
tomography: inferring the flow-free photon inverse slope $T_{\gamma0}$ from
the dilepton inverse slope $T_{\ell\bar\ell}$.  To isolate this thermometer
step, thermal emission only ($\alpha_{\rm pre}=0$) is considered.  The
radial-flow strength is fixed at $\kappa_r=1$, while
$T_{0,\mathrm{peak}}$ is varied from $0.24$ to $0.48~\mathrm{GeV}$ with
the geometry held unchanged.  For each
temperature history, $T_{\ell\bar\ell}$ is extracted from the flowing
dilepton spectrum, while $T_{\gamma0}$ is obtained from the corresponding
photon calculation with transverse flow removed.  The two inverse slopes
therefore probe the same thermal evolution, with $T_{\gamma0}$ providing
the flow-free photon reference to be inferred from $T_{\ell\bar\ell}$.

\begin{figure}[t]
    \centering
    \includegraphics[width=0.95\linewidth]
    {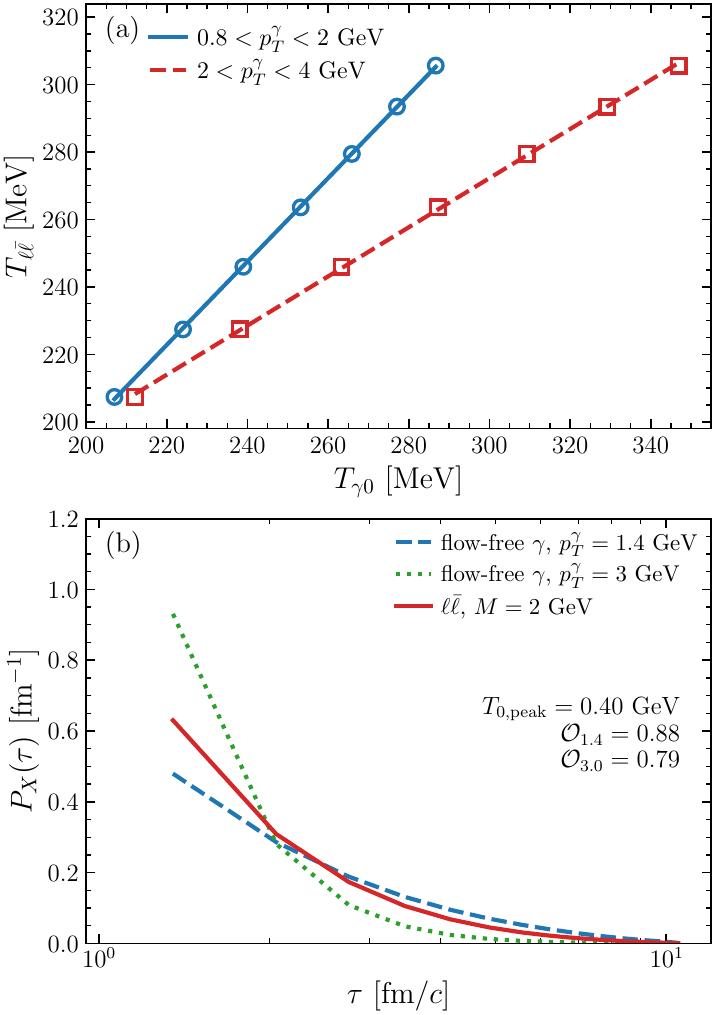}
    \caption{
    (a) Dilepton inverse slope $T_{\ell\bar\ell}$ in
    $1<M<3~\mathrm{GeV}$ versus the flow-free photon inverse slope
    $T_{\gamma0}$ extracted in two photon momentum windows as
    $T_{0,\mathrm{peak}}$ is varied at fixed geometry.
    (b) Normalized proper-time emission profiles at
    $T_{0,\mathrm{peak}}=0.40~\mathrm{GeV}$ for flow-free photons at
    $p_T^\gamma=1.4$ and $3~\mathrm{GeV}$ and for $M=2~\mathrm{GeV}$
    dileptons from the baseline $\kappa_r=1$ calculation.
    }
    \label{fig:thermometer}
\end{figure}

The resulting thermometer relations are shown in
Fig.~\ref{fig:thermometer}(a).  For the default photon window
$0.8<p_T^\gamma<2~\mathrm{GeV}$, the relation is
\begin{equation}
T_{\ell\bar\ell}
=
-49.4~\mathrm{MeV}
+
1.237\,T_{\gamma0}.
\label{eq:thermal-thermometer}
\end{equation}
The RMS deviation from this linear relation is about
$0.35~\mathrm{MeV}$,\footnote{Unless stated otherwise, an RMS deviation
denotes $\left[N^{-1}\sum_i(\Delta_i)^2\right]^{1/2}$ over the
corresponding scan points, where $\Delta_i$ is the residual from the
relation or reference being tested.}
with a maximum deviation of about $0.6~\mathrm{MeV}$.  A similarly tight
relation is obtained when $T_{\gamma0}$ is extracted from
$2<p_T^\gamma<4~\mathrm{GeV}$, for which the RMS deviation is about
$0.6~\mathrm{MeV}$, although the relation itself shifts appreciably.
At fixed $T_{\ell\bar\ell}$, the harder photon window gives a larger
$T_{\gamma0}$, reflecting the greater sensitivity of higher-$p_T$ photons
to the earlier, hotter part of the evolution
\cite{Shen:2013vja,Paquet:2022wgu}.  Thus the thermometer relation remains highly
precise in both photon windows, while its numerical form depends on the
kinematic selection.

The unusually tight relation can be understood from how the two inverse
slopes respond to the underlying thermal evolution.  Comparing
$T_{\gamma0}$ and $T_{\ell\bar\ell}$ with the weighted thermal-stage
initial temperature $\langle T_{\rm in}\rangle$ used in
Ref.~\cite{Du:2026aux}, neither inverse slope is itself perfectly linear
in this thermal scale.  Their nonlinear deviations,
however, are almost perfectly correlated, with a Pearson correlation
coefficient $\rho\simeq0.998$. This common response is shown explicitly in
Appendix~\ref{app:common-thermal-scale}. The common nonlinear response therefore
largely cancels when $T_{\ell\bar\ell}$ is plotted directly against
$T_{\gamma0}$, producing the nearly linear thermometer relation seen in
Fig.~\ref{fig:thermometer}(a).

This behavior is consistent with the picture of dilepton thermometry
developed in Ref.~\cite{Du:2026aux}, where $T_{\ell\bar\ell}$ was shown
to track an emission-weighted thermal scale that changes systematically
with the underlying temperature evolution.  The present result shows that
the flow-free photon inverse slope responds to the same evolution in nearly
the same way.  The precision of the
$T_{\ell\bar\ell}$--$T_{\gamma0}$ mapping can therefore be understood as
a consequence of their closely matched response to the common thermal
history.

The shift of the thermometer relation with the photon momentum window can
be understood from how the different observables sample the evolution in
time.  Figure~\ref{fig:thermometer}(b) compares their normalized
proper-time emission profiles.  For a selected photon momentum or dilepton mass, the normalized emission
profile is defined as
\begin{equation}
P_X(\tau)
=
\frac{1}{Y_X}\frac{dY_X}{d\tau},
\qquad
\int d\tau\,P_X(\tau)=1,
\label{eq:time-profile}
\end{equation}
where $X=\gamma$ or $\ell\bar\ell$ and $Y_X$ is the corresponding yield at
the selected kinematics.  The profiles are characterized by their mean
emission times,
$\langle\tau\rangle_X=\int d\tau\,\tau P_X(\tau)$,
and by their normalized overlap,
\begin{equation}
{\cal O}_{\gamma,\ell\bar\ell}
=
\int d\tau\,
\min\!\left[
P_\gamma(\tau),P_{\ell\bar\ell}(\tau)
\right],
\label{eq:emission-overlap}
\end{equation}
which ranges from zero for nonoverlapping profiles to unity for identical
ones.

For the representative choice
$T_{0,\mathrm{peak}}=0.40~\mathrm{GeV}$, the
$p_T^\gamma=1.4~\mathrm{GeV}$ photons have
$\langle\tau\rangle_\gamma\simeq2.9~\mathrm{fm}/c$, compared with
$\langle\tau\rangle_{\ell\bar\ell}\simeq2.4~\mathrm{fm}/c$ for
$M=2~\mathrm{GeV}$ dileptons, with an overlap
${\cal O}_{\gamma,\ell\bar\ell}\simeq0.88$.  The
$p_T^\gamma=3~\mathrm{GeV}$ photons are emitted substantially earlier,
with $\langle\tau\rangle_\gamma\simeq1.8~\mathrm{fm}/c$, and their overlap
with the dilepton profile decreases to about $0.79$.  The harder photon
selection therefore places greater weight on the earlier, hotter part of
the thermal evolution, consistent with the larger $T_{\gamma0}$ in
Fig.~\ref{fig:thermometer}(a).

These emission histories explain why the numerical thermometer relation
changes with the photon momentum window.  Harder photons weight earlier and
hotter emission more strongly, changing the response of $T_{\gamma0}$ to
the same thermal evolution and therefore shifting its mapping to
$T_{\ell\bar\ell}$.  This window dependence is distinct from the origin of
the near-linearity itself: the common thermal-scale response controls the
precision of the mapping, whereas the different emission histories control
its numerical form.

The analysis above isolates the thermal contribution and varies the thermal
scale at fixed geometry.  More realistic hydrodynamic calculations in
Ref.~\cite{Du:2025dot}, in which centrality and beam energy also change,
likewise found an approximately common
$T_{\ell\bar\ell}$--$T_{\gamma0}$ relation.  This suggests that the mapping
can remain robust under broader variations of the thermal evolution,
although its numerical form is not universal.  A similarly tight relation
is also found when a prethermal component is included, and this more general
case is incorporated into the flow reconstruction below.  Before examining
the source dependence in detail, the thermometer inference is first bypassed
to ask what physical flow quantity is represented by $v_r^{\rm eff}$ when
$T_{\gamma0}$ is known directly.

\subsection{Emission-weighted Doppler response}

The second step of the tomography construction is isolated next by asking what
physical flow quantity is encoded by the photon Doppler response
$v_r^{\rm eff}$.  To avoid mixing this question with uncertainties from the
thermometer inference, the directly calculated $T_{\gamma0}$ is used rather
than the dilepton-inferred $\widehat T_{\gamma0}$.  The electromagnetic
source is fixed to $\alpha_{\rm pre}=1$ and
$\tau_{\rm chem}=1~\mathrm{fm}/c$, including both thermal and prethermal
radiation, while varying $T_{0,\mathrm{peak}}=0.24$--$0.48~\mathrm{GeV}$
and $\kappa_r=0.5,\ 1,$ and $1.5$ independently.  For each case,
$T_\gamma$ is extracted from the total photon spectrum and $T_{\gamma0}$
from the same temperature and source history with transverse flow removed.
The resulting $v_r^{\rm eff}$ therefore isolates the flow-induced spectral
response without invoking the thermometer reconstruction.

A natural first question is whether this spectral response simply follows a
bulk-like measure of radial expansion \cite{Du:2024pbd,Du:2025dot}.  As a reference, a bulk-like average
is defined as
\begin{equation}
\langle v_r\rangle
=
\frac{\sum_s\int_s d^4x\,[T_s^{\rm eff}]^4\gamma v_r}
     {\sum_s\int_s d^4x\,[T_s^{\rm eff}]^4\gamma},
\qquad s\in\{{\rm th,pre}\},
\label{eq:bulkflow}
\end{equation}
where $\gamma=\sqrt{1+u_\perp^2}$ is the local Lorentz factor.  For the
thermal stage, $T_{\rm th}^{\rm eff}=T$ and $T^4$ is proportional to the
energy density for the conformal equation of state.  In the prethermal
stage, $[T_{\rm pre}^{\rm eff}]^4$ serves as the analogous bulk-like weight,
since $T_{\rm pre}^{\rm eff}$ is an emission scale rather than an
equilibrium thermodynamic temperature.  Equation~\eqref{eq:bulkflow}
therefore provides an energy-density-like, Lorentz-factor-weighted
spacetime average of the radial velocity.  It is used here as a bulk-like
reference rather than as a unique definition of the fireball radial flow.

The photon spectral response, however, is sensitive not only to the
magnitude of the local radial motion but also to its direction relative to
the observed photon.  Suppressing the longitudinal dependence,
\begin{equation}
p\cdot u
\sim
\gamma p_T
\left(1-v_r\cos\Delta\phi\right),
\end{equation}
so flow aligned with the photon momentum lowers the photon energy in the
local rest frame and enhances the emission at fixed $p_T$ through the
approximately Boltzmann-like dependence
$\exp[-p\cdot u/T_s^{\rm eff}]$.  The velocity component relevant for the
Doppler response is therefore naturally weighted by the same local emission
that builds the photon spectrum.  This motivates the rate-weighted
projected radial velocity
\begin{equation}
\left\langle v_r\cos\Delta\phi\right\rangle_R
=
\frac{\sum_s\int_s d^4x\,
R_{\gamma,s}\,v_r\cos\Delta\phi}
{\sum_s\int_s d^4x\,R_{\gamma,s}},
\label{eq:projectedflow}
\end{equation}
where $s\in\{{\rm th,pre}\}$,
$\Delta\phi=\phi_p-\phi_s$ is the angle between the photon momentum and the
local radial velocity, and $R_{\gamma,s}$ denotes the local photon-emission
weight entering the spectrum in stage $s$.  The source azimuth $\phi_s$ is
already included in $d^4x$; by azimuthal symmetry the photon direction
$\phi_p$ may be chosen arbitrarily.

\begin{figure}[t]
    \centering
    \includegraphics[width=0.95\linewidth]
    {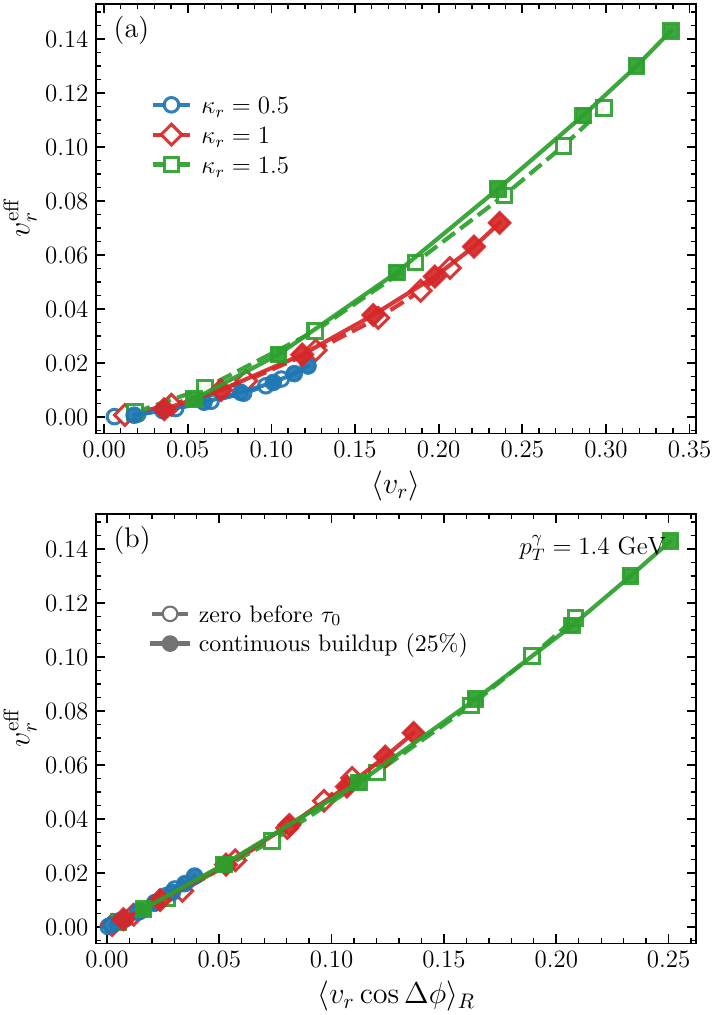}
    \caption{
    Direct effective spectral flow $v_r^{\rm eff}$ for the fixed
    thermal-plus-prethermal source versus
    (a) the bulk-like radial velocity $\langle v_r\rangle$ and
    (b) the rate-weighted projected radial velocity
    $\langle v_r\cos\Delta\phi\rangle_R$, evaluated at
    $p_T^\gamma=1.4~\mathrm{GeV}$.
    Curves connect increasing $T_{0,\mathrm{peak}}$ at fixed $\kappa_r$.
    Dashed curves with open symbols use the zero-before-$\tau_0$ velocity
    history, while solid curves with filled symbols use the
    continuous-buildup history.
    }
    \label{fig:flowmapping}
\end{figure}

For the comparison in Fig.~\ref{fig:flowmapping},
Eq.~\eqref{eq:projectedflow} is evaluated at
$p_T^\gamma=1.4~\mathrm{GeV}$, near the
center of the default $0.8<p_T^\gamma<2~\mathrm{GeV}$ photon fit window.
The zero-before-$\tau_0$ velocity history, shown by the dashed curves with
open symbols, is considered first.
Figure~\ref{fig:flowmapping}(a) compares the direct $v_r^{\rm eff}$ with
the bulk-like average $\langle v_r\rangle$.  Increasing
$T_{0,\mathrm{peak}}$ at fixed $\kappa_r$ traces a smooth sequence, but the
sequences for different $\kappa_r$ remain systematically separated and do
not collapse onto a single tight linear trend.  Comparable values of
$\langle v_r\rangle$ can therefore correspond to different
$v_r^{\rm eff}$, showing that the bulk-like radial motion does not uniquely
determine the photon spectral response.

A much tighter relation appears in Fig.~\ref{fig:flowmapping}(b) when the
same results are plotted against
$\langle v_r\cos\Delta\phi\rangle_R$.  The different $\kappa_r$ sequences
collapse much more closely onto a common trend across the
zero-before-$\tau_0$ scan.  The photon Doppler response is therefore more
directly associated with the
emission-weighted velocity component along the observed photon momentum
than with the overall radial speed.

The role of the rate weighting in this tighter correspondence can be seen
most clearly in the small-flow limit.  Without the emission weight, the
projection $v_r\cos\Delta\phi$ averages to zero in an azimuthally symmetric
collision.  The photon emission rate, however, is itself modified by the
same directional Doppler factor.  Writing schematically
\[
R_\gamma
\simeq
R_0
\left[
1+a\,v_r\cos\Delta\phi+\cdots
\right],
\]
and inserting this expansion into Eq.~\eqref{eq:projectedflow}, the
zeroth-order contribution to the numerator vanishes upon azimuthal
integration, while the leading surviving term is proportional to
$v_r^2\cos^2\Delta\phi$.  The azimuthally integrated photon spectrum has
the same symmetry: its term linear in $v_r\cos\Delta\phi$ cancels, so the
leading flow-induced spectral modification is likewise quadratic in the
characteristic flow amplitude \cite{Paquet:2022wgu}.  The fitted inverse
slope therefore satisfies
$T_\gamma-T_{\gamma0}=O(v_r^2)$ in the small-flow limit.  Since
Eq.~\eqref{eq:veff} gives
$v_r^{\rm eff}\simeq
(T_\gamma-T_{\gamma0})/T_{\gamma0}$
for a small slope difference, $v_r^{\rm eff}$ also begins at quadratic
order.  The rate-weighted projection and the effective spectral flow
therefore have the same leading dependence on the underlying radial-flow
field, helping to explain their much tighter correspondence.

The continuous-buildup history, in which radial flow begins developing
already at $\tau_{\rm init}$, provides a further test of this
interpretation and is shown by the filled markers connected by solid lines
in Fig.~\ref{fig:flowmapping}.  This variation changes the development of
radial motion while leaving the temperature and electromagnetic source
histories fixed.  The scan extends to larger $v_r^{\rm eff}$, with the
maximum increasing from about $0.11$ to $0.14$, and the relation with the
bulk-like $\langle v_r\rangle$ shifts visibly.  By contrast, the points
remain close to the common trend in
$\langle v_r\cos\Delta\phi\rangle_R$.  The tighter correspondence with
the projected velocity therefore persists and is not specific to the
zero-before-$\tau_0$ prescription.

The Doppler response also depends on the photon momentum window.  At higher
$p_T$, the flow-induced modification of the fitted photon inverse slope is
substantially reduced and can change sign, yielding a slightly negative
direct $v_r^{\rm eff}$, consistent with Ref.~\cite{Paquet:2022wgu}.  The direct $v_r^{\rm eff}$ is therefore most
naturally interpreted as a finite-window, rate-weighted directional Doppler
response rather than as a unique bulk radial velocity, although it remains
closely correlated with the overall radial expansion.  The experimentally
relevant construction, however, does not know $T_{\gamma0}$ directly and
must infer it from dileptons.  This inference step is restored next to
examine how the reconstruction changes when the electromagnetic source
differs from the one used to establish the thermometer relation.

\subsection{Source-transfer bias and apparent flow}

The previous subsection deliberately used the directly calculated
$T_{\gamma0}$ to isolate the photon Doppler response.  The first step of the
experimentally relevant construction is now restored, with the flow-free
photon reference inferred from the dilepton inverse slope, to examine how
reliably the thermometer relation can be transferred from a model calculation
to the physical system.  This is important because the
thermometer relation must be established within a model, whose
electromagnetic source may omit or imperfectly describe contributions
present in the measured spectra.  A mismatch between the modeled and
physical source composition can therefore bias the inferred
$T_{\gamma0}$ and, consequently, the reconstructed flow.  The omission of
prethermal radiation is used as a controlled example of such a mismatch
between the modeled and physical electromagnetic source.

The source-transfer bias in the reconstructed photon spectral flow is first
isolated through a zero-direct-flow test at
$T_{0,\mathrm{peak}}=0.36~\mathrm{GeV}$.  The underlying evolution uses
the baseline $\kappa_r=1$ radial-flow history, but transverse flow is
deliberately removed when calculating the photon spectrum.  The photon
inverse slope $T_\gamma$ is therefore extracted directly from a flow-free
spectrum, so that $T_\gamma=T_{\gamma0}$ and the direct
$v_r^{\rm eff}$ vanishes by construction.  The dilepton spectrum retains
the baseline $\kappa_r=1$ thermal-stage flow used in constructing the
thermometer relation; its residual flow sensitivity from the finite
pair-momentum acceptance is sub-MeV in the present setup.  The prethermal
strength $\alpha_{\rm pre}$ is varied for
$\tau_{\rm chem}=0.3,\ 1,$ and $3~\mathrm{fm}/c$;
$T_\gamma$ and $T_{\ell\bar\ell}$ are extracted from the resulting
spectra, and $\widehat T_{\gamma0}$ is then deliberately inferred using
the \emph{thermal-only} thermometer relation of
Eq.~\eqref{eq:thermal-thermometer}.  Combining this inferred reference
with $T_\gamma$ through Eq.~\eqref{eq:reconstructed-veff} gives
$\widehat v_r^{\rm eff}$.

This construction provides a null test of the reconstruction.  Because the
photon spectrum is evaluated with transverse flow removed,
$T_\gamma=T_{\gamma0}$ and the direct $v_r^{\rm eff}$ is exactly zero.
If the thermal-only thermometer relation remained valid after the source
composition changed, it would also give
$\widehat T_{\gamma0}=T_\gamma$ and hence
$\widehat v_r^{\rm eff}=0$.  A source mismatch can instead shift the photon
and dilepton inverse slopes away from the correlated response encoded by
the thermal-only relation, causing the inferred $\widehat T_{\gamma0}$ to
fall below or rise above $T_\gamma$.  The reconstruction then yields a
nonzero $\widehat v_r^{\rm eff}$ even though the direct photon spectral
flow vanishes.  Such a reconstruction-induced signal is termed
\emph{apparent flow}; its sign is positive or negative depending on whether
$\widehat T_{\gamma0}$ lies below or above $T_\gamma$, respectively.

\begin{figure}[t]
    \centering
    \includegraphics[width=0.95\linewidth]
    {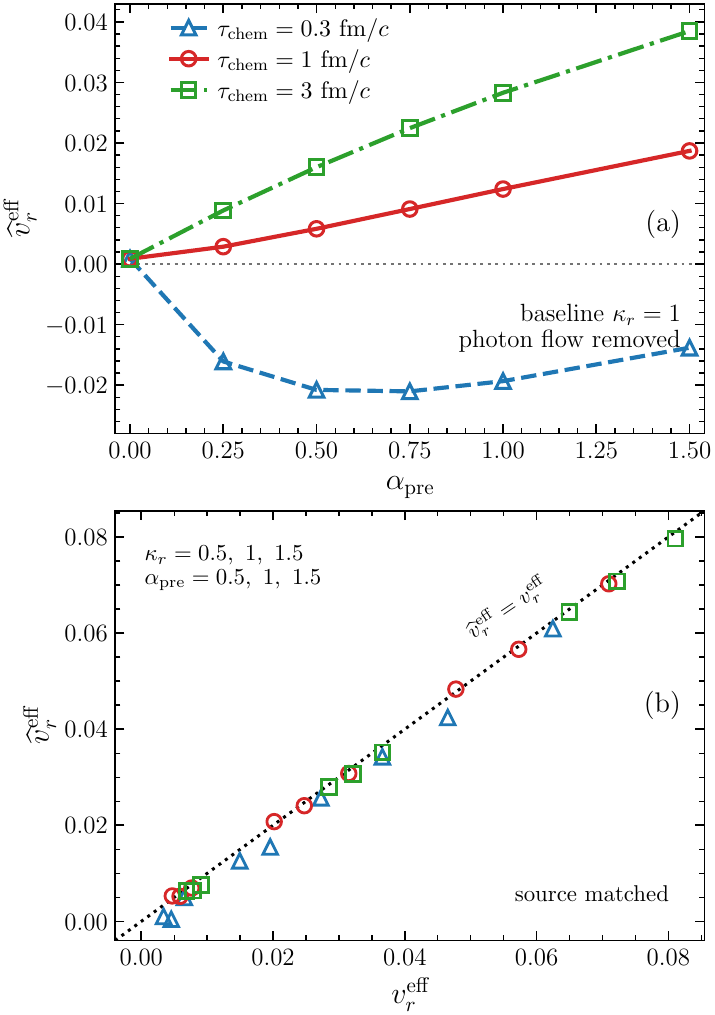}
    \caption{
    (a) Apparent flow $\widehat v_r^{\rm eff}$ in the zero-direct-flow
    source-transfer test at
    $T_{0,\mathrm{peak}}=0.36~\mathrm{GeV}$, shown versus
    $\alpha_{\rm pre}$ for three chemical-equilibration times
    $\tau_{\rm chem}$.  The photon spectrum is calculated without transverse
    flow, so the direct $v_r^{\rm eff}$ vanishes, while the flow-free photon
    reference is inferred using the thermal-only thermometer relation of
    Eq.~\eqref{eq:thermal-thermometer}.  Any nonzero
    $\widehat v_r^{\rm eff}$ therefore arises from source mismatch.
    (b) Closure of the full reconstruction using a separate thermometer
    relation for each $(\alpha_{\rm pre},\tau_{\rm chem})$ source
    prescription.  The reconstructed effective spectral flow
    $\widehat v_r^{\rm eff}$ is compared with the directly calculated
    $v_r^{\rm eff}$ for
    $\kappa_r=0.5,\ 1,$ and $1.5$,
    $\alpha_{\rm pre}=0.5,\ 1,$ and $1.5$, and
    $\tau_{\rm chem}=0.3,\ 1,$ and $3~\mathrm{fm}/c$.
    The dotted diagonal denotes exact closure,
    $\widehat v_r^{\rm eff}=v_r^{\rm eff}$.
    }
    \label{fig:source-transfer}
\end{figure}

Figure~\ref{fig:source-transfer}(a) shows that the transferred thermometer
relation can generate apparent flow of either sign.  At
$\alpha_{\rm pre}=0$ the source is thermal only and matches the relation,
so the reconstruction closes near zero.  As the prethermal contribution
increases, slow chemical equilibration produces an increasingly positive
$\widehat v_r^{\rm eff}$, whereas fast equilibration initially produces
negative apparent flow whose magnitude is largest at intermediate
$\alpha_{\rm pre}$ before decreasing again.  The bias therefore depends
not only on the amount of prethermal radiation, but on how the additional
source changes the photon and dilepton inverse slopes relative to the
thermal-only thermometer relation.

The different behaviors originate from the unequal quark-fugacity
dependence of the two channels.  The prethermal stage samples an earlier
and hotter part of the evolution and therefore tends to increase both
$T_\gamma$ and $T_{\ell\bar\ell}$.  For slow chemical equilibration,
however, the $\lambda_q^2$ weighting suppresses the early dilepton
contribution more strongly than the $\lambda_q$ weighting of the photon
rate.  As $\alpha_{\rm pre}$ increases, $T_\gamma$ therefore rises more rapidly
than the dilepton-implied flow-free photon reference, so that
$T_\gamma>\widehat T_{\gamma0}$ and hence
$\widehat v_r^{\rm eff}>0$.
The departure from the thermal-only correlation consequently grows as the
prethermal contribution becomes stronger.

For fast chemical equilibration, $\lambda_q$ approaches unity early and the
difference between the photon and dilepton fugacity weightings becomes much
smaller.  At intermediate $\alpha_{\rm pre}$, the prethermal contribution can
instead raise $T_{\ell\bar\ell}$ enough that the thermal-only mapping gives
$\widehat T_{\gamma0}>T_\gamma$, and therefore
$\widehat v_r^{\rm eff}<0$.
As the prethermal component becomes increasingly important in both
channels, their correlated response can move back toward the thermal-only
thermometer relation.\footnote{In the fast-equilibration limit,
$\lambda_q\simeq1$ through most of the prethermal interval, so the early
source approaches a chemically equilibrated extension of the same
Bjorken-like cooling trajectory used for the thermal stage.  Although the
resulting spectra sample an earlier and hotter temperature range, their
correlated photon--dilepton response can therefore resemble that of a
thermal-only evolution at a higher thermal scale.}
The inferred $\widehat T_{\gamma0}$ then moves back toward $T_\gamma$,
reducing the magnitude of the negative apparent flow and producing the
turnover in Fig.~\ref{fig:source-transfer}(a).  The nonmonotonic behavior
thus reflects the changing relative response of the two electromagnetic
channels rather than radial-flow dynamics.

The source-transfer bias does not imply that the photon--dilepton method
fails to reconstruct the effective spectral flow when the electromagnetic
source is modeled consistently.  To test this directly, a separate
thermometer relation is established for each source prescription
$(\alpha_{\rm pre},\tau_{\rm chem})$.  For each prescription,
$T_{0,\mathrm{peak}}$ is scanned from $0.24$ to $0.48~\mathrm{GeV}$ in
steps of
$0.04~\mathrm{GeV}$ and, as in Fig.~\ref{fig:thermometer}, pair the
dilepton inverse slope from the baseline flowing calculation
($\kappa_r=1$) with the corresponding flow-free photon inverse slope
obtained after removing transverse flow.  The electromagnetic source
prescription is otherwise held fixed.

This source-matched thermometer relation is then applied at
$T_{0,\mathrm{peak}}=0.36~\mathrm{GeV}$ to infer
$\widehat T_{\gamma0}$ from the baseline dilepton inverse slope and combine
it with photon spectra calculated at
$\kappa_r=0.5,\ 1,$ and $1.5$.  The analysis considers
$\alpha_{\rm pre}=0.5,\ 1,$ and $1.5$ and
$\tau_{\rm chem}=0.3,\ 1,$ and $3~\mathrm{fm}/c$, giving the 27
combinations shown in Fig.~\ref{fig:source-transfer}(b).  For the
zero-before-$\tau_0$ velocity history used here, the prethermal contribution
remains flow free, while varying $\kappa_r$ changes the thermal-stage
radial flow.

The reconstructed effective spectral flow $\widehat v_r^{\rm eff}$ is
then compared with the direct $v_r^{\rm eff}$ obtained using the calculated
$T_{\gamma0}$.  The points lie close to the exact-closure line
$\widehat v_r^{\rm eff}=v_r^{\rm eff}$, with an RMS deviation of
$0.002$ and a maximum absolute deviation of $0.004$.  Thus, when the
thermometer relation and the analyzed spectra use a consistent
electromagnetic source prescription, the reconstructed effective spectral
flow closely reproduces the direct result.  This closure confirms that the
much larger apparent flow in Fig.~\ref{fig:source-transfer}(a) arises
primarily from transferring a thermometer relation across mismatched source
compositions.

This closure also sharpens the practical limitation.  In an experimental
application, the early electromagnetic source is not known in advance, so
the appropriate source-matched thermometer relation cannot simply be
assumed.  The relevant question is therefore whether source variations
leave distinguishable signatures in the measured spectra themselves.  The
next question is whether combining several photon momentum and dilepton mass
windows provides enough independent information to separate such source
variations from genuine radial flow.

\subsection{Multiwindow diagnosis of source ambiguity}

The source-transfer test above exposes an ambiguity in the reconstructed
effective flow.  A nonzero $\widehat v_r^{\rm eff}$ can arise from genuine
radial flow, which changes the observed photon inverse slope relative to its
flow-free value, but it can also be generated or biased by a mismatch in the
electromagnetic source composition, which shifts the inferred flow-free
reference $\widehat T_{\gamma0}$.  A single photon--dilepton spectral pair
therefore does not by itself determine how much of the reconstructed signal
originates from radial motion and how much from the source assumption.

This ambiguity need not be irreducible, because flow and source variations
affect different parts of the photon and dilepton spectra differently.  In
particular, different photon momentum windows sample different stages of the
evolution and exhibit different Doppler responses, while different dilepton
mass windows have different sensitivities to thermal and prethermal
radiation \cite{Paquet:2022wgu,Du:2026aux}.  The question is therefore whether several
spectral windows provide sufficiently distinct response patterns to help
separate radial flow from variations of the early electromagnetic source.

The original two-slope construction is compared with an extended
four-window set.  The original observables are the photon inverse slope in
$0.8<p_T^\gamma<2~\mathrm{GeV}$ and the dilepton inverse slope in
$1<M<3~\mathrm{GeV}$.  The extended set uses photon inverse slopes in
$0.8<p_T^\gamma<2$ and $2<p_T^\gamma<4~\mathrm{GeV}$ together with
dilepton inverse slopes in $1<M<2$ and $2<M<3~\mathrm{GeV}$.  Denoting
these observables by $O_a$, their local fractional responses to
$\boldsymbol{\theta}=(\kappa_r,\alpha_{\rm pre},\tau_{\rm chem})$ are characterized through the dimensionless response matrix
\begin{equation}
J_{ai}
=
\frac{\theta_i}{O_a}
\frac{\partial O_a}{\partial\theta_i}
=
\frac{\partial\ln O_a}{\partial\ln\theta_i}.
\label{eq:response-matrix}
\end{equation}
Thus each matrix element measures the fractional change of an observable
produced by a fractional change of a parameter.  This normalization removes
the trivial dependence on the units and absolute scales of the different
observables and parameters, allowing their responses to be compared on the
same relative basis.

Each column of $J$ can then be viewed geometrically as a response vector in
the space of the selected spectral slopes.  If two such response vectors are
proportional, the corresponding parameter changes alter the observables in
the same relative pattern and cannot be distinguished locally.  The
singular-value decomposition of $J$ provides a compact way
to determine how many independent first-order response patterns are present
and how strongly they are expressed.\footnote{Writing
$J=U\Sigma V^{T}$, the columns of $V$ are the right-singular vectors,
which specify combinations of parameter variations, while the columns of
$U$ give the associated response patterns in observable space.}
The nonnegative singular values are denoted by $\sigma_i(J)$ and ordered as
$\sigma_1\geq\sigma_2\geq\cdots$.  The number of nonzero singular values
gives the number of independent local response directions, while a small
$\sigma_i(J)$ identifies a direction to which the selected observables
respond only weakly.  The corresponding right-singular vector shows which
combination of $\kappa_r$, $\alpha_{\rm pre}$, and $\tau_{\rm chem}$
generates that direction.  The relevant question is therefore not simply
whether more spectral slopes are measured, but whether the additional
windows introduce a genuinely new and sufficiently distinct response
pattern.

This local analysis uses the zero-before-$\tau_0$ velocity history and
evaluates the response around
$(\kappa_r,\alpha_{\rm pre},\tau_{\rm chem})
=(1,1,1~\mathrm{fm}/c)$ for
$T_{0,\mathrm{peak}}=0.28$, $0.36$, and $0.44~\mathrm{GeV}$.  Here
$T_{0,\mathrm{peak}}$ is treated as an external thermal-scale coordinate
rather than as an additional parameter in $J$; separate response matrices
are evaluated at the three representative temperatures.  The derivatives
are evaluated by centered finite differences using endpoint pairs
$(0.5,1.5)$, $(0.75,1.25)$, and
$(0.8,1.2)~\mathrm{fm}/c$ for
$\kappa_r$, $\alpha_{\rm pre}$, and $\tau_{\rm chem}$, respectively.
To verify that the inferred response rank is not an artifact of the chosen
finite-difference steps, all three half-widths are varied simultaneously
between $0.5$ and $1.5$ times these values.  For the four-window set, rank
three is preserved at every tested temperature, while the weakest singular
value changes by only a few percent.  The purpose of this analysis is to
diagnose the local structure of the model response rather than to perform
an experimental parameter extraction.

\begin{figure}[t]
    \centering
    \includegraphics[width=0.95\linewidth]
    {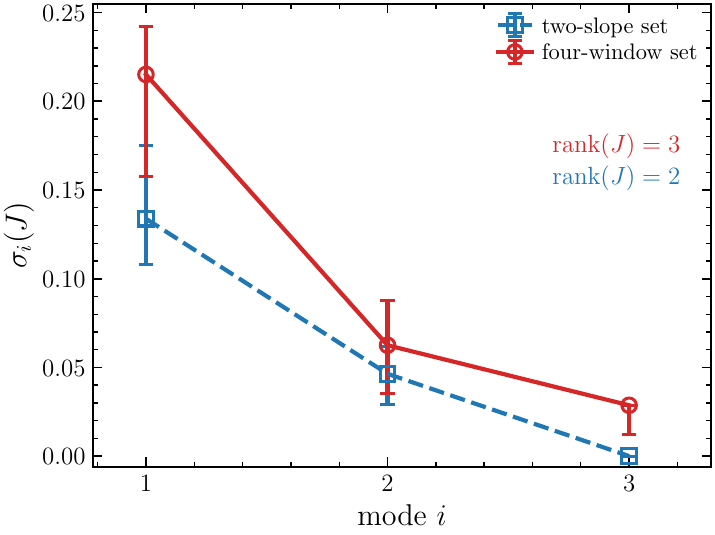}
    \caption{
    Singular values $\sigma_i(J)$ of the local fractional-response matrix
    $J$ for the original two-slope and extended four-window observable
    sets.  The response is evaluated with the zero-before-$\tau_0$ velocity
    history around
    $(\kappa_r,\alpha_{\rm pre},\tau_{\rm chem})
    =(1,1,1~\mathrm{fm}/c)$.  Points show the singular values at
    $T_{0,\mathrm{peak}}=0.36~\mathrm{GeV}$; the bars span the minimum and
    maximum values obtained at
    $T_{0,\mathrm{peak}}=0.28$, $0.36$, and $0.44~\mathrm{GeV}$ and do not
    represent statistical uncertainties.
    }
    \label{fig:multiwindow-response}
\end{figure}

With only the original photon and dilepton slopes, the $2\times3$ response
matrix can contain at most two independent response directions.  At least
one local combination of
$\kappa_r$, $\alpha_{\rm pre}$, and $\tau_{\rm chem}$ is therefore
indistinguishable at first order.  The important question is whether the
additional spectral windows merely repeat these two responses or introduce
a genuinely new one.  Figure~\ref{fig:multiwindow-response} shows the
latter: the four-window set has a nonzero third singular value at the
tested temperatures.  The additional windows therefore introduce a third
independent first-order response pattern within the controlled model,
rather than simply providing redundant measurements of the leading two.

The third response direction is nevertheless considerably weaker than the
leading two.  The small value of $\sigma_3(J)$ means that there exists a
particular combination of parameter variations to which the four spectral
slopes respond only weakly.  To determine which parameters dominate this
weak direction, inspection of the corresponding right-singular vector shows
that it is dominated by $\alpha_{\rm pre}$ and $\tau_{\rm chem}$, with only a small
$\kappa_r$ component.  The remaining weakly distinguishable direction
therefore lies primarily within the prethermal-source sector: an
appropriate combined variation of the prethermal strength and
chemical-equilibration timescale produces only a small net response across
the selected spectral windows.  By contrast, $\kappa_r$ contributes little
to this weak direction, indicating that radial-flow variations are more
readily distinguished from variations of the early electromagnetic source
within the present model.

This analysis is local and does not establish global parameter uniqueness
or include experimental uncertainties and covariance; the singular values
should therefore be interpreted as diagnostics of the model response rather
than as parameter constraints.  Its main implication is that the source
mismatch responsible for the apparent flow in
Fig.~\ref{fig:source-transfer}(a) is not spectrally featureless.  Source
variations leave correlated changes across photon momentum and dilepton mass
windows that differ from the pattern generated by radial flow.  The same
finite-window dependence that limits a single-window reconstruction can
therefore become useful diagnostic information when several windows are
considered together.  Multiwindow measurements can, in principle, test the
electromagnetic-source assumptions entering photon--dilepton tomography
rather than requiring those assumptions to be fixed a priori.

\section{Summary and outlook}

This study has separated and tested the two ingredients underlying
photon--dilepton radial-flow tomography.  The first is the thermometer
relation that uses the dilepton invariant-mass inverse slope
$T_{\ell\bar\ell}$ to infer the flow-free photon inverse slope
$T_{\gamma0}$.  For thermal emission, this relation is highly precise:
the two inverse slopes respond to changes of the thermal evolution in
nearly the same way, even though neither is itself a simple measure of a
single temperature.  Its numerical form nevertheless depends on the
spectral selection, because different photon momentum windows weight
different parts of the emission history.  More realistic hydrodynamic
results indicate that the relation can remain robust when the underlying
evolution is varied more broadly, although its coefficients are not
universal.

The second ingredient is the conversion of the flow-induced modification
of the photon inverse slope into an effective spectral flow.  When
$T_{\gamma0}$ is known directly, $v_r^{\rm eff}$ remains closely
correlated with the overall radial expansion but tracks the
photon-emission-weighted projected velocity
$\langle v_r\cos\Delta\phi\rangle_R$ considerably more closely than the
bulk-like average $\langle v_r\rangle$.  This correspondence persists when
the assumed radial-flow history is changed.  The effective spectral flow is
therefore most naturally interpreted as a finite-window,
emission-weighted directional Doppler response rather than as a unique
bulk radial velocity.

A leading controlled limitation appears when the thermometer relation is
transferred between different electromagnetic source compositions.
Prethermal radiation changes the photon and dilepton inverse slopes
differently through their distinct quark-fugacity dependence, so applying
a thermal-only thermometer relation to spectra containing an additional
prethermal component can bias the inferred $\widehat T_{\gamma0}$.  In the
zero-direct-flow null test, this mismatch generates nonzero
$\widehat v_r^{\rm eff}$ even though the direct photon spectral flow
$v_r^{\rm eff}$ vanishes, a reconstruction-induced signal termed here
apparent flow.  By contrast,
when the thermometer relation is established using the same source
prescription as the spectra being analyzed, the reconstructed
$\widehat v_r^{\rm eff}$ closely reproduces the direct
$v_r^{\rm eff}$ over the tested finite-flow cases.  The bias is therefore
primarily a source-transfer effect rather than an intrinsic failure of the
photon--dilepton reconstruction.

The source dependence need not remain an uncontrolled ambiguity.  Different
photon momentum and dilepton mass windows respond differently to radial
flow and to changes of the early electromagnetic source.  In the local
response analysis, extending the original two-slope construction to two
photon and two dilepton windows introduces a third independent first-order
response direction, allowing radial flow to be distinguished more
effectively from variations of the prethermal-source sector.  The weakest
direction remains dominated by the prethermal strength
$\alpha_{\rm pre}$ and chemical-equilibration timescale
$\tau_{\rm chem}$, indicating that these two source properties are more
difficult to separate from one another than from radial flow.  Thus the
same finite-window dependence that limits a single-window reconstruction
can become useful diagnostic information when several spectral windows are
considered together.

For experiment, this suggests a practical extension of the original
construction.  Rather than relying on a single photon momentum window and
a single dilepton mass window, inverse slopes can be extracted in several
kinematic regions and tested simultaneously.  A common source description
that reproduces the corresponding photon and dilepton responses would
support the inferred flow signal, whereas systematic inconsistencies among
the windows would indicate sensitivity to the assumed electromagnetic
source.  The different windows can therefore be used not only to improve
the radial-flow reconstruction but also to test and constrain the source
model entering it.  Importantly, this does not require the measured
radiation to be separated experimentally into pre-equilibrium, QGP, and
hadronic components; the relevant requirement is that their combined
contribution to the measured spectra be described consistently.

A quantitative application will require the controlled ingredients studied
here to be embedded in a more realistic dynamical and electromagnetic
description.  In particular, pre-equilibrium, thermal QGP, and hadronic
radiation should be propagated consistently through a common evolution,
rather than treating the source composition only through the restricted
variation used here.  The same framework should eventually incorporate
realistic viscous hydrodynamics and the corresponding nonequilibrium
corrections to the electromagnetic emission rates.  Shear- and
bulk-viscous corrections, chemical nonequilibrium, and, at lower collision
energies, finite-density effects can all modify the spectral shapes from
which the inverse slopes are extracted.  Nonthermal contributions that
remain in the relevant kinematic regions must likewise be controlled at
the level required by the reconstruction.  These effects need not be
disentangled one by one in the measured spectra, but their combined impact
on the photon--dilepton relation must be modeled consistently.

This perspective may become particularly relevant across the Beam Energy
Scan and toward lower collision energies
\cite{Bzdak:2019pkr,Du:2024wjm,NA60:2022sze,An:2021wof}.  Dilepton
measurements already span RHIC Beam Energy Scan energies and lower-energy
baryon-rich systems
\cite{STAR:2015zal,STAR:2023wta,HADES:2019auv,STAR:2024bpc}.  In this regime, the finite
nuclear-overlap and heating stage become increasingly important for the early
evolution \cite{Goes-Hirayama:2025lhs,Wu:2025iix}, while realistic
$(3+1)$D calculations exhibit strongly baryon-rich and longitudinally
structured dynamics \cite{Du:2023efk,Du:2023gnv,Du:2019obx}.  These features, together with the
changing role of finite-density electromagnetic radiation, further
complicate the photon and dilepton emission history
\cite{Du:2024pbd}.  Chemical nonequilibrium can introduce an
additional source dependence beyond the controlled setting considered
here, while the development of radial flow during these early stages is
itself of considerable interest.  Extending photon--dilepton tomography to
realistic $(3+1)$D baryon-rich dynamics with finite-density emission rates
would therefore test both sides of the construction: how robustly
dileptons determine the flow-free photon reference as the source changes,
and what part of the evolving radial motion is encoded by the photon
Doppler response.  Measurements in several photon momentum and dilepton
mass windows could then provide complementary information on radial
expansion and electromagnetic-source dynamics, allowing the two to be
constrained together rather than requiring either to be fixed in advance.

\section*{Acknowledgements}
The author acknowledges helpful conversations with Ulrich Heinz. This work was supported in part by the U.S. Department of Energy, Office of Science,
Office of Nuclear Physics, under Grant No.~DE-AC02-05CH11231. The author acknowledges the use of ChatGPT for grammar refinement, clarity enhancement, and analysis code optimization.


\appendix*

\renewcommand{\thefigure}{A\arabic{figure}}
\setcounter{figure}{0} 
\setcounter{equation}{0} 
\renewcommand{\thetable}{\arabic{table}}


\appendix
\section{Common thermal-scale response of photon and dilepton inverse slopes}
\label{app:common-thermal-scale}

The main text attributes the unusually precise
$T_{\ell\bar\ell}$--$T_{\gamma0}$ relation to the closely matched response
of the two inverse slopes to the underlying thermal evolution.  This
response is shown explicitly here.  The same thermal-only,
fixed-geometry scan as in Fig.~\ref{fig:thermometer} is used, with
$T_{0,\mathrm{peak}}=0.24$--$0.48~\mathrm{GeV}$ varied while the other
model inputs are kept fixed.  The dilepton inverse slope
$T_{\ell\bar\ell}$ is extracted in the default $1<M<3~\mathrm{GeV}$ mass
window from the baseline $\kappa_r=1$ thermal calculation, while
$T_{\gamma0}$ is extracted in the default
$0.8<p_T^\gamma<2~\mathrm{GeV}$ photon window from the same temperature
and source history with transverse flow removed from the photon
calculation.

\begin{figure}[t]
    \centering
    \includegraphics[width=0.95\linewidth]
    {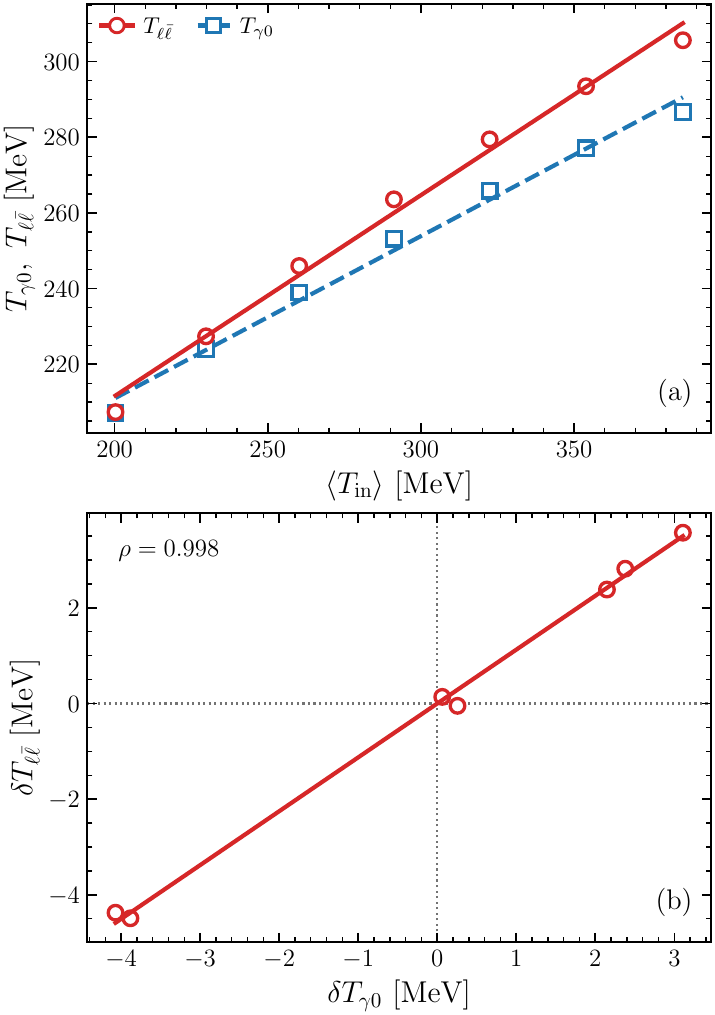}
    \caption{
    (a) Flow-free photon inverse slope $T_{\gamma0}$ and dilepton inverse
    slope $T_{\ell\bar\ell}$ versus the weighted thermal-stage initial
    temperature $\langle T_{\rm in}\rangle$ for the thermal-only,
    fixed-geometry scan used in the main text.  Lines show separate linear
    fits to the two inverse slopes.
    (b) Residual dilepton inverse slope
    $\delta T_{\ell\bar\ell}$ versus the corresponding photon residual
    $\delta T_{\gamma0}$ after subtracting their respective best-fit linear
    dependences on $\langle T_{\rm in}\rangle$.
    }
    \label{fig:common-thermal-scale}
\end{figure}

The initialization is characterized by the $e\gamma$-weighted thermal-stage
initial temperature used in Ref.~\cite{Du:2026aux},
\begin{equation}
\left\langle T_{\rm in}\right\rangle
=
\frac{
\int d^2x_\perp\,
T(\tau_0,\mathbf{x}_\perp)
e[T(\tau_0,\mathbf{x}_\perp)]
\gamma(\tau_0,\mathbf{x}_\perp)
}{
\int d^2x_\perp\,
e[T(\tau_0,\mathbf{x}_\perp)]
\gamma(\tau_0,\mathbf{x}_\perp)
}.
\label{eq:appendix-initial-temperature}
\end{equation}
For the present initialization $\gamma=1$.
Figure~\ref{fig:common-thermal-scale}(a) shows that both inverse slopes
increase approximately linearly with the same initial thermal scale, but
neither relation is perfectly linear.  Centering the fits at the mean
sampled value
$\langle T_{\rm in}\rangle=291.9~\mathrm{MeV}$, the corresponding
best-fit linear responses are defined as
\begin{align}
T_{\gamma0}^{\rm lin}
\bigl(\langle T_{\rm in}\rangle\bigr)
&=
250.4~\mathrm{MeV}
+
0.429
\left(
\langle T_{\rm in}\rangle-291.9~\mathrm{MeV}
\right),
\label{eq:appendix-photon-linear-fit}
\\
T_{\ell\bar\ell}^{\rm lin}
\bigl(\langle T_{\rm in}\rangle\bigr)
&=
260.4~\mathrm{MeV}
+
0.531
\left(
\langle T_{\rm in}\rangle-291.9~\mathrm{MeV}
\right).
\label{eq:appendix-dilepton-linear-fit}
\end{align}
The RMS residuals are
$2.72$ and $3.06~\mathrm{MeV}$ for
$T_{\gamma0}$ and $T_{\ell\bar\ell}$, respectively.
Thus the two inverse slopes have different overall sensitivities to the
thermal scale, while both exhibit small departures from a purely linear
response.  The strong correlation of $T_{\gamma0}$ with the initial
thermal scale is also consistent with the sensitivity of thermal-photon
spectra to the early temperature evolution discussed in
Ref.~\cite{Paquet:2022wgu}.

The deviations of the two sets of points from their respective linear fits
also follow a strikingly similar pattern in
Fig.~\ref{fig:common-thermal-scale}(a).  To quantify this common nonlinear
response, the residuals relative to
Eqs.~\eqref{eq:appendix-photon-linear-fit} and
\eqref{eq:appendix-dilepton-linear-fit} are defined as
\begin{equation}
\delta T_{\gamma0}
=
T_{\gamma0}
-
T_{\gamma0}^{\rm lin}
\bigl(\langle T_{\rm in}\rangle\bigr),
\qquad
\delta T_{\ell\bar\ell}
=
T_{\ell\bar\ell}
-
T_{\ell\bar\ell}^{\rm lin}
\bigl(\langle T_{\rm in}\rangle\bigr).
\label{eq:appendix-thermal-residuals}
\end{equation}
As shown in Fig.~\ref{fig:common-thermal-scale}(b), these residuals are
almost perfectly correlated,
\begin{equation}
\rho\!\left(
\delta T_{\gamma0},
\delta T_{\ell\bar\ell}
\right)
=
0.998,
\end{equation}
and are related approximately by
$\delta T_{\ell\bar\ell}
\simeq
1.13\,\delta T_{\gamma0}$.
The RMS scatter about this residual--residual relation is only
$0.17~\mathrm{MeV}$.  Thus the departures of the two inverse slopes from
their respective linear thermal-scale responses are not independent: as
the thermal history is varied, they deviate from the linear trends in
nearly the same way.

This common nonlinear response explains why it largely cancels when
$T_{\ell\bar\ell}$ is plotted directly against $T_{\gamma0}$, as discussed
in the main text.  The separate linear
$T_{\gamma0}$--$\langle T_{\rm in}\rangle$ and
$T_{\ell\bar\ell}$--$\langle T_{\rm in}\rangle$ relations have RMS
residuals of $2.72$ and $3.06~\mathrm{MeV}$, whereas the direct
$T_{\ell\bar\ell}$--$T_{\gamma0}$ relation in
Fig.~\ref{fig:thermometer}(a) has an RMS residual of only
$0.35~\mathrm{MeV}$, with a maximum deviation of about
$0.6~\mathrm{MeV}$.  The precision of the photon--dilepton thermometer
relation therefore does not require either inverse slope to be an exactly
linear thermometer of $\langle T_{\rm in}\rangle$.  Rather, it follows
from their closely matched response, including their common nonlinear
dependence, on the same thermal evolution.

\bibliography{master_refs}
\end{document}